\documentclass[final]{IEEEtran}

\usepackage[colorlinks]{hyperref}
\usepackage{url}
\hypersetup{urlcolor={blue}}
\usepackage{amsmath}
\usepackage{algorithm}
\usepackage[noend]{algpseudocode}
\usepackage{graphicx}
\usepackage{cleveref}
\usepackage{amssymb}
\usepackage{eqnarray}
\usepackage{subcaption}
\usepackage{booktabs}
\usepackage[margin={.75in,.75in}]{geometry}
\makeatletter
\usepackage{etoolbox}
\makeatletter

\title{Microsimulation Analysis for Network Traffic Assignment (MANTA) at Metropolitan-Scale for Agile Transportation Planning}

\author{
\large{Pavan Yedavalli}\\
\texttt{University of California, Berkeley}\\
\texttt{\textcolor{black}{pavyedav@berkeley.edu}}\\
\large{Krishna Kumar} \\
\texttt{University of Texas at Austin}\\
\texttt{\textcolor{black}{krishnak@utexas.edu}}\\
\large{Paul Waddell} \\
\texttt{University of California, Berkeley}\\
\texttt{\textcolor{black}{waddell@berkeley.edu}}\\
}

\begin{document}

\maketitle 

\begin{abstract}

Abrupt changes in the environment, such as unforeseen events due to climate change, have triggered massive and precipitous changes in human mobility. The ability to quickly predict traffic patterns in different scenarios has become more urgent to support short-term operations and long-term transportation planning. This requires modeling entire metropolitan areas to recognize the upstream and downstream effects on the network. However, there is a well-known trade-off between increasing the level of detail of a model and decreasing computational performance. To achieve the level of detail required for traffic microsimulation, current implementations often compromise by simulating small spatial scales, and those that operate at larger scales often require access to expensive high performance computing systems or have computation times on the order of days or weeks that discourage productive research and real-time planning. This paper addresses the performance shortcomings by introducing a new platform, MANTA (Microsimulation Analysis for Network Traffic Assignment), for traffic microsimulation at the metropolitan-scale. MANTA employs a highly parallelized GPU implementation that is capable of running metropolitan-scale simulations within a few minutes. The runtime to simulate all morning trips, using half-second timesteps, for the nine-county San Francisco Bay Area is just over four minutes, not including routing and initialization. This computational performance significantly improves the state of the art in large-scale traffic microsimulation. MANTA expands the capacity to analyze detailed travel patterns and travel choices of individuals for infrastructure planning.
\end{abstract}

\section{Introduction}
Rapid global urbanization and an increase in the frequency of extreme events, such as climate change-induced disruptive weather occurrences, are forcing us to re-examine the way we design and improve the resilience of cities, including their transportation infrastructure. Transportation simulation models offer the ability to perform sensitivity analyses and ex-ante evaluation of the impact of potential infrastructure investments~\cite{kokkinogenisNextgenerationTrafficSimulation2011, garcia-doradoDesigningLargescaleInteractive2014, waddellUrbanSimModelingUrban2002}. These simulations explore human mobility patterns, which are motivated by the need to engage in mandatory and discretionary activities. They are carried out on various modes, including walking, biking, driving, or TNC services. The dynamics of traffic flow are affected by factors such as frequency of trips, vehicle occupancy, length of the journeys, route choices, and driving speeds, producing congestion, traffic emissions, and an increase in traffic accidents~\cite{zhaoAgentBasedModelABM2019}. In addition, certain transportation simulation, such as emergency evacuation planning of a city in an extreme weather event, requires near real-time transportation planning. Hence, to address the need for metropolitan-scale emergency scenarios and broader infrastructure planning by policymakers and urban planners, we develop a fast metropolitan-scale traffic simulation engine capable of characterizing individual behaviors.

Traffic modelers use three alternative types of traffic assignment models to predict the impact of travel demand on the network: (1) macroscopic, (2) mesoscopic, and (3) microscopic, in decreasing order of traveler aggregation and increasing order of granularity~\cite{milamClosingInducedVehicle2017, waraichPerformanceImprovementsLargeScale2015}. Macroscopic models are based on the continuum assumption in classical fluid mechanics. The traffic flow is treated as continuous, similar to a flow of a liquid in a pipe, rather than that comprising of discrete vehicles~\cite{maerivoetTransportationPlanningTraffic2005a}.  These macroscopic models are useful in analyzing traffic systems covering a wide area, often across regions, and on highways where the overall speed dictates the macroscopic behavior~\cite{kokkinogenisNextgenerationTrafficSimulation2011}. Unlike macroscopic models that assume a continuous vehicular flow on a road link (edge), mesoscopic models employ aggregated volume delay functions, by clustering a set of vehicles into packets and evaluating the movement of these clusters~\cite{kokkinogenisNextgenerationTrafficSimulation2011}. In contrast to these models, microscopic traffic simulation models provide even greater granularity, giving explicit consideration to the interactions between individual vehicles within a traffic stream and employing characteristics such as vehicle lengths, speeds, accelerations, time, and space headways~\cite{toledoMicroscopicTrafficSimulation2005}. 

Metropolitan-scale transportation modeling has been dominated by the macroscopic and mesoscopic models, due to their relative computational efficiency and familiarity~\cite{kotusevskiReviewTrafficSimulation2009}. However, one of the significant drawbacks of these simulators is their lack of granularity. They are limited by the accuracy of representing real-world vehicle dynamics, especially in congested regimes and for emergency scenarios~\cite{toledoMicroscopicTrafficSimulation2005, axhausenActivityBasedApproaches1992}. Traffic flow dynamics are naturally an outcome of the interaction of a many-vehicle system, where each vehicle exhibits different characteristics~\cite{yangMicroscopicTrafficSimulator1996}. Only a microsimulation model can capture these intricacies of individual components and complex interactions with reasonable accuracy~\cite{loderUnderstandingTrafficCapacity2019, yangMicroscopicTrafficSimulator1996, toledoMicroscopicTrafficSimulation2005, geroliminisIdentificationAnalysisQueue2011}. However, microsimulation has a high computational cost due to the granularity required in simulating the vehicle movements. Hence, metropolitan-scale microsimulation has generally been impractical~\cite{saidallahComparativeStudyUrban2016, kokkinogenisNextgenerationTrafficSimulation2011}. Although many traffic simulators exist, such as MATSim, SUMO, AIMSUN, Polaris, TRANSSIM, VISSIM, and DynaMIT, among others, these simulators are not designed to tackle metropolitan-scale traffic microsimulation efficiently~\cite{Horni2016, dlr46740, barceloDynamicNetworkSimulation2005, auldPOLARISAgentbasedModeling2016, saxenaProblemSolvingUncertainty2016, parkMicroscopicSimulationModel2003, ben-akivaDynaMITSimulationbasedSystem1998}. As a result, techniques such as sampling a small fraction of the transportation demand are currently employed to achieve metropolitan-scale traffic simulation in a reasonable amount of time and computational cost.

This paper introduces a massively parallelized GPU implementation of a metropolitan-scale microsimulation engine - Microsimulation Analysis for Network Traffic Assignment (MANTA). MANTA is an agile metropolitan-scale microsimulator capable of efficiently simulating over 7 million agents at a spatial scale as large as the San Francisco (SF) Bay Area, in under 10 minutes. First, we present the components of the simulation, then the mathematical theory and implementation of the simulator, followed by the results of a case study in the Bay Area. We then present the calibration and validation of the simulator, performance benchmarks, limitations and future work, and finally the conclusions.

\section{Components}
The objective of this study is to perform a metropolitan-scale microsimulation of vehicular traffic of the SF Bay Area, incorporating individual trips on a typical workday morning. The microsimulator builds on the initial implementation by~\cite{garcia-doradoDesigningLargescaleInteractive2014, waddellIntegratedPipelineArchitecture2018}. In this section, the network generation, demand creation, routing, and simulation architectures are described in detail. 


\subsection{Street Network}

\begin{figure}
    \centering
    \includegraphics[width=.45\textwidth]{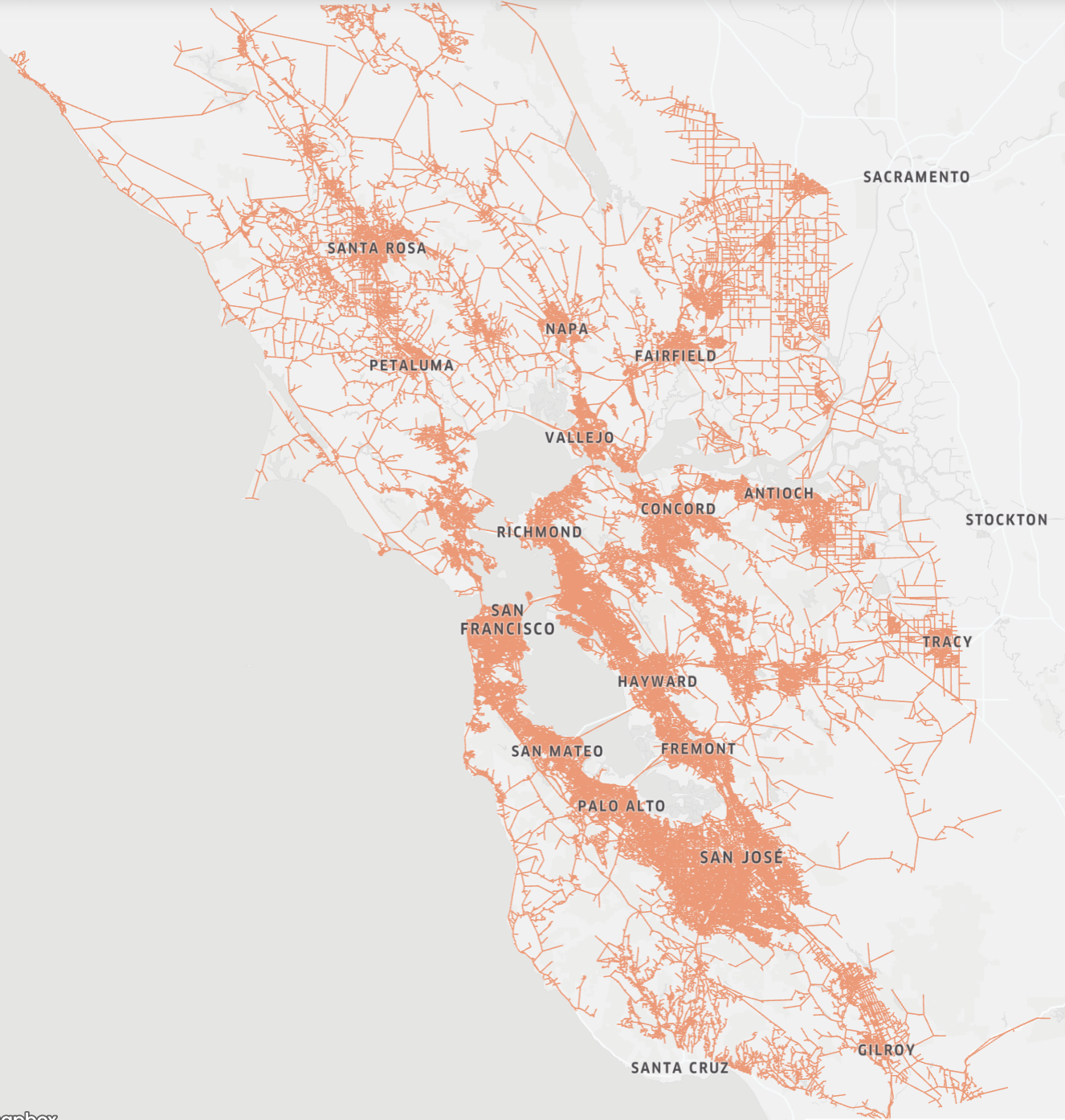}
    \caption{The Bay Area Network with 224,223 nodes and 549,008 edges}
    \label{fig:bay_area_edges}
\end{figure}

For the case study, we use the SF Bay Area, which includes nine counties. The street network is constructed from the OpenStreetMap (OSM) network within the polygonal hull of the counties in the metropolitan area using the OSMnx library~\cite{boeingOSMnxNewMethods2017a}. The network contains all roads in the SF Bay Area, from large primary roads to tertiary streets. The OSM network includes intermediary points representing curves or bends in the road links, which are topologically unnecessary for network analysis~\cite{waddellIntegratedPipelineArchitecture2018, boeingOSMnxNewMethods2017a}. As these intermediary nodes do not represent a real intersection, they are removed from the network used for the microsimulation. The final network used for the microsimulation only includes nodes that represent physical intersections or dead-ends. The resulting network is a connected graph of the San Francisco Bay Area, where there exists a path from any node on the graph to any other node in the graph. There are no hanging nodes without a path.~\Cref{fig:bay_area_edges} shows the full network of the SF Bay Area with 224,223 nodes and 549,008 edges. The number of lanes, length, and free-flow speeds for each edge are extracted from OSM data or imputed. The speed limit of each edge is taken from OSM, if available; if not, a free-flow speed limit is computed based on the number of lanes and the type of road. If the number of lanes is not available, then a recommended default value from OSM is used, depending on the type of road. For instance, a tertiary road without a specified number of lanes is given a default speed limit of 20 mph, and a motorway without a specified number of lanes is given a default speed limit of 57.5 mph.


\subsection{Demand}

The origin-destination (OD) demand is derived from data generated by the Bay Area Metropolitan Transportation Commission (MTC) travel model~\cite{metropolitantransportationcommissionPlanBayArea2017}. For simplicity and considering the near-symmetric bimodality of morning and evening travel demand~\cite{metropolitantransportationcommissionPlanBayArea2017}, this study only considers the morning trips between 5 AM and 12 PM. The demand is further restricted to only automobile trips, which includes private automobiles, transportation network company (TNC) vehicles, and driving trips to transit stops. This demand does not consider public transit trips, such as buses, or large freight trucks, as the Bay Area MTC travel demand model is synthesized from only household travel data. While the data do not accurately reflect the exact real-world network congestion, leaving approximately 250K trips off the SF Bay Area road network for the 5 AM to 12 PM timeframe, the travel demand data remain comprehensive with approximately 3 million trips ~\cite{metropolitantransportationcommissionPlanBayArea2017}

The OD pairs constructed from the MTC model are available at the granularity of traffic analysis zones (TAZ). TAZs are population density-based geographical areas typically larger than blocks but smaller than zip codes. The SF Bay Area consists of 1454 TAZs. The travel demand data is at the TAZ level (i.e., the OD pair represents a trip from an origin TAZ to a destination TAZ). The microsimulation requires an origin and destination node within its respective TAZ, which is achieved using a two-step process. First, all the nodes in the road network are mapped to their respective TAZ polygon. Second, for each TAZ-level OD pair, the origin-destination nodes are determined by sampling from a uniform distribution within the origin and destination TAZs, respectively. This process differs from~\cite{waddellIntegratedPipelineArchitecture2018}, where origins and destinations are assigned to the centroid of their respective TAZs rather than being distributed across different nodes within the TAZ. The random assignment avoids unrealistic congestion at the centroids of the TAZs. The final OD demand has 3,269,864 travelers.


\subsection{Routing}
After generating the network and the corresponding OD demand, the next step of the simulation is routing, where we compute the shortest-path between each origin and destination pair. Routing algorithms have been bottlenecks in many traffic models, requiring either significant pre-processing time or great computational cost~\cite{dellingEngineeringRoutePlanning2009a}. The edge lengths, obtained from OpenStreetMap, represent the length of the road link in meters, which is used as the weight of the edge in the graph network for calculating the shortest path between nodes.


One of the significant contributions of this paper is the integration of a parallelized Dijkstra's priority queue single-source shortest-path (SSSP) algorithm, described in \cite{zhaoAgentBasedModelABM2019}, in which only the OD pairs required in the simulation are computed. The priority-queue algorithm is parallelized with a hybrid MPI/OpenMP scheme, which allows for linear scaling with millions of agents on a cluster. An open addressing scheme-based hashmap is used to store key-value pairs of edge weights, hence updating the edge weights during the simulation and computing the shortest-path becomes more efficient. This open addressing scheme improves the performance of hashmaps by 20\%, providing quicker access to edges and connectivity. A simulation with 3.2 million OD pairs routes on the large SF Bay Area network is calculated within 62 minutes on an Intel I9 processor with 2 threads per core and 14 cores per socket.

\subsection{Microsimulation}

\begin{figure}
    \centering
    \includegraphics[width=.45\textwidth]{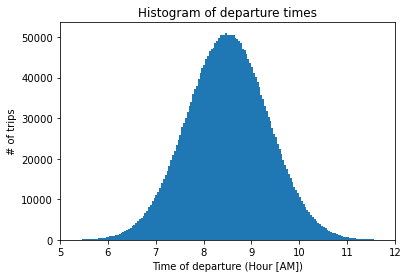}
    \caption{Departure times are chosen between 5 AM to 12 PM to model the morning hours. It follows a Gaussian distribution in which the bulk of the trips begin between 6:30 AM and 8:30 AM.}
    \label{fig:dep_times}
\end{figure}

The microsimulation framework we adopt is an enhanced and extended version of the architecture developed by~\cite{garcia-doradoDesigningLargescaleInteractive2014}. The vehicles move in discrete timesteps of $\delta t = .5$ seconds, following the state of the art microsimulators today~\cite{dowlingTrafficAnalysisToolbox2004}. The simulation described in this paper models a typical morning workday from 5 AM to 12 PM. Each traveler in the OD demand is randomly assigned a departure time within this specified range by sampling from a normal distribution that roughly mimics the morning peak-hour behavior, with a peak around 7:30 AM and a standard deviation of 45 minutes. The departure time specified for individual vehicles in the simulation is presented in~\Cref{fig:dep_times}.

At each timestep, the vehicle's travel time, position, and velocity are updated. MANTA employs a unique traffic atlas concept, akin to a texture atlas in the computer graphics community or a discretization step in signal processing. Each road segment of length $l$ is discretized into $\frac{l}{t_m}$ different compartments, where $t_m$ is the length of each compartment, in meters, and is specified to be 1 meter in this implementation.  Each compartment of the road is assigned to a specific byte in the computer memory. Hence, a road represented in the computer memory is a contiguous sequence of bytes. Each byte in memory can be occupied by at most one vehicle, and records the velocity of the vehicle and reflects their real position on the road lane~\cite{garcia-doradoDesigningLargescaleInteractive2014}. Each road segment is represented as a new row in the 2D traffic atlas. The 2D array layout of the road network requires less memory, making it parallelizable on a GPU rather than requiring a graph-based approach. The traffic atlas significantly reduces the computational cost of finding nearby vehicles, as it only involves looking up the status of neighboring cells in the memory array. The lookup scheme on a 2D grid to identify the vehicle's speed and its neighbors' speeds is parallelized with each thread on a GPU querying a specific block of memory address, which significantly speeds up the computation on the GPU. This approach varies from the traditional method, which requires checking the entire edge for neighboring vehicles.

In MANTA, the vehicular movement on an edge is dictated by conventional car following, lane changing, and gap acceptance algorithms \cite{toledoMicroscopicTrafficSimulation2005}. The well-known Intelligent Driver Model (IDM), as shown in~\Cref{eq:idm}, is used to control the vehicle dynamics through the network \cite{treiberTrafficFlowDynamics2013}.
\begin{equation} 
    \dot{v} = a (1 - (\frac{v}{v_o})^\delta - (\frac{s_o + Tv + \frac{v\Delta v}{2\sqrt{ab}}}{s})^2)
    \label{eq:idm}
\end{equation}
where $\dot{v}$ is the current acceleration of the vehicle, $a$ is the acceleration potential of the vehicle, $v$ is the current speed of the vehicle, $v_o$ is the speed limit of the edge, $\delta$ is the acceleration exponent, $s$ is the gap between the vehicle and the leading vehicle, $s_0$ is the minimum spacing allowed between vehicles when they are at a standstill, $T$ is the desired time headway, and $b$ is the braking deceleration of the vehicle \cite{garcia-doradoDesigningLargescaleInteractive2014, waddellIntegratedPipelineArchitecture2018}. The exact position of each vehicle at a given timestep is computed by double integrating the acceleration value $\dot{v}$. Table~\ref{Tab:idm_params} shows the range of possible values for the IDM parameters $a$, $b$, $T$, and $s_0$, which are derived based on the simulations from~\cite{treiberTrafficFlowDynamics2013}. These parameters are calibrated for the microsimulation of the SF Bay Area using real-world data, with the process described in~\cref{sec:calibration}.
\begin{table}   
    \centering
    \begin{tabular}{ l  c  c }
        \toprule
         Parameter & Value & Units \\
         \midrule
         $a$ & $\mathcal{N}(1,10)$ & $\frac{m}{s^2}$ \\  
         $b$ & $\mathcal{N}(1,10)$ & $\frac{m}{s^2}$ \\
         $T$ & $\mathcal{N}(.1,2)$ & $s$ \\
         $s_0$ & $\mathcal{N}(1,5)$ & $m$\\
         \bottomrule
    \end{tabular}
    \caption{IDM parameter ranges, derived from
    ~\cite{treiberTrafficFlowDynamics2013}}
    \label{Tab:idm_params}
\end{table}

MANTA is designed to be modular to incorporate different vehicle profiles and dynamics. In the future, these could include models for adaptive cruise control (ACC) and autonomous vehicles. Specifically, since the IDM leveraged by MANTA has been used in ACC systems in previous literature, it can be quickly adapted to accommodate for ACC in the future~\cite{kimDesignAdaptiveCruise2012}. In ACC, the IDM maintains an appropriate relative distance to the lead vehicle, in contrast to standard cruise control systems, whose objective is to maintain only a target speed~\cite{kimDesignAdaptiveCruise2012, milanesModelingCooperativeAutonomous2014}. Automated and platooned systems are also capable of cooperative adaptive cruise control (CACC), where there is communication among interacting vehicles~\cite{kimDesignAdaptiveCruise2012}. Since the control system in each vehicle can retrieve information from the adjacent vehicles, such as the acceleration, velocity, and braking, these values can be stored as the IDM parameters in MANTA. The current setup uses the same values across all vehicles, based on calibration with real-world data (see~\cref{sec:calibration}).

In addition to car following, vehicles can also change lanes within an edge. There are two types of lane changes: mandatory and discretionary~\cite{garcia-doradoDesigningLargescaleInteractive2014}. Mandatory lane changes occur when the vehicle must take an exit off the road, while discretionary lane changes occur during overtaking or voluntary movements~\cite{garcia-doradoDesigningLargescaleInteractive2014}. The lane changing model gives the vehicle an exponential probability from switching from a discretionary lane change to a mandatory lane change, as shown in~\Cref{eq:lane_changing}.
\begin{equation}
    m_i = 
    \begin{cases} 
        e^{-(x_i - x_0)^2} & x_i \geq x_0 \\
        1 & x_i \leq x_0
        \label{eq:lane_changing}
    \end{cases}
\end{equation}
where $m_i$ is the probability of a mandatory lane change for vehicle $i$, $x_i$ is the distance of vehicle $i$ to an exit or intersection, and $x_0$ is the distance of a critical location, which may be the position of a particular message sign (such as a final exit warning) to the exit or intersection (statically set to 1)~\cite{iqbalDevelopmentOriginDestination2014a, yangMicroscopicTrafficSimulator1996}. Intuitively, as the vehicle travels further along in a path, its probability of making a lane change to make a turn or exit increases. 

Once a vehicle has decided to change lanes, the maneuver is performed if the lead and lag gaps are acceptable. A lane change dynamic involves interaction among three vehicles: the merging vehicle $i$, the lead vehicle $a$, and the lag vehicle $b$. The critical lead or lag gap for a successful lane change is defined as the minimum distance to the following or lagging vehicle at which a lane change can be performed, respectively, as shown in~\Cref{eq:gap1} and~\Cref{eq:gap2}.
\begin{align}
    \label{eq:gap1}
    g_{lead} &= \max(g_a, g_a + \alpha_{a}v_i + \alpha_{i}(v_i - v_a)) + \epsilon_a  \\ 
    \label{eq:gap2}
    g_{lag} &= \max(g_b, g_b + \alpha_{i}v_i + \alpha_{b}(v_i - v_b)) + \epsilon_b
\end{align}
where $g_{lead}$ is the critical lead gap for a lane change, $g_{lag}$ is
the critical lag gap for a lane change, $g_a$ is the desired lead gap for a lane change, $g_b$ is the desired lag gap for a lane change, $v_i$ is the speed of the merging vehicle $i$, $v_a$ is the speed of the lead vehicle, $v_b$ is the speed of the lag vehicle, $\alpha_{i}$ is the anticipation time of vehicle $i$ attempting to change lanes in between vehicles $a$ and $b$ (in seconds), which have anticipation times $\alpha_a$ and $\alpha_b$, respectively. Because drivers perceive distances and times differently, this anticipation time varies from vehicle to vehicle. The values of $\alpha_i$, $\alpha_a$, and $\alpha_b$ were chosen in the range [0.05,0.40], based on historically calibrated models~\cite{choudhuryModelingCooperativeLanechanging2007}). If $\alpha_{i}$, $\alpha_a$, or $\alpha_b$ equals 0, this means the anticipation time of vehicle $i$, $a$, or $b$, for vehicle $i$ to make a lane change is 0 (i.e., the vehicle anticipates an extremely aggressive lane change with a low desired gap). Importantly, the anticipated gap is calculated based on the assumption that other drivers will maintain their current accelerations. For example, if the lag vehicle of the merging driver is decelerating, the anticipated gap will increase. Finally, $\epsilon_a$ and $\epsilon_b$ are the random components, each normally distributed with mean 0 and standard deviation 1, with units in meters. The values of $g_a$ and $g_b$ are a function of the speeds of the merging, leading, and lagging vehicles ($i, a, b$, respectively), but typically range between 1 to 5 meters at speeds below 25 mph and 5 to 10 meters at speeds above 25 mph~\cite{garcia-doradoDesigningLargescaleInteractive2014}.

The representation and modeling of intersections in this initial application of the traffic simulator is simplistic and not representative of diverse real-world dynamics at intersections. We consider two different types of traffic control. Case 1 traffic control is a flashing red light at each node, where only one vehicle can move into the intersection at a particular time. If the node contains $n$ inbound edges and $m$ outbound edges, the system will create a round-robin of the $nm$ combinations for all cars to pass through the intersection based on their position in their lane queue~\cite{waddellIntegratedPipelineArchitecture2018}. Case 2 traffic control assigns every node as a green light, where all cars pass through the intersection with no delay. This is clearly not realistic for most nodes that have stop signs or traffic lights, but is plausible for nodes along highway interchanges. However, congestion on the edge itself is still modeled, as each edge has a finite capacity for vehicles and will not allow new vehicles from adjacent edges if the capacity is reached. The results of the simulations for these two cases are discussed in~\cref{sec:lights}.


\section{OD and Routing Results}
\begin{figure} 
    \centering
    \includegraphics[width=.45\textwidth]{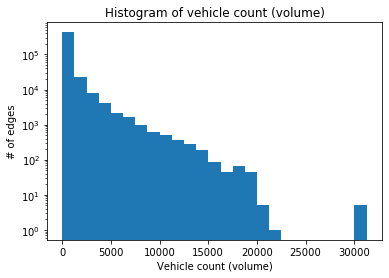}
    \caption{Histogram (log y-axis) showing the number of edges that see a particular vehicle count across the time range simulated. The Bay Bridge is the most heavily traversed link in the Bay Area, as it connects two major economic hubs: San Francisco and Oakland. The subset of edges representing the Bay Bridge and its necessary feeder edges (7 in total) sees a significant traffic volume of 30K trips, which has approximately 7.5K vehicles more than the next highest edges. Most edges see fewer than 100 vehicles in the timeframe.}
    \label{fig:volume}
\end{figure}

\begin{figure} 
    \includegraphics[width=.5\textwidth]{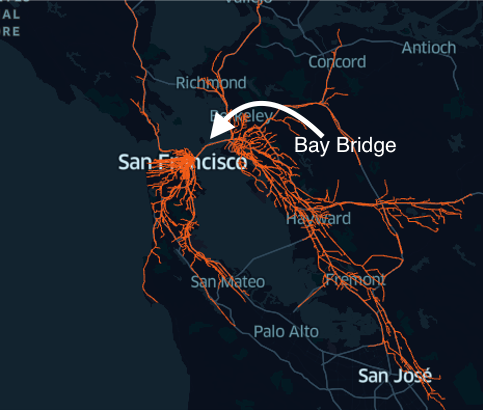}
    \caption{A network showing all possible routes through the Bay Bridge}
    \label{fig:transbay}
\end{figure}

Preliminary travel patterns already emerge from the initial routing calculation.~\Cref{fig:volume} displays a log histogram of edge volumes across the network, showing that the bulk of edges have vehicle counts below 1000 and are traversed infrequently. Only a small fraction of the edges account for the majority of the most frequently traversed routes.
Routes across the Bay Bridge are shown in~\Cref{fig:transbay}. Unsurprisingly, the Bay Bridge remains a unique outlier, as it accounts for a maximum volume of 31270 vehicles in the seven-hour duration. From ~\cite{actransitBayBridgeCorridor2010} by AC Transit and ARUP, 41727 trips out of a total of approximately 4M trips traverse the Bay Bridge between 5 AM and 12 PM, representing 1\% of all trips. This proportion of Bay Bridge traversals matches the proportion from the routing output at roughly 0.98\% (31270 trips out of 3.2M total trips in the SF Bay Area).



\section{Simulation Results}

Infrastructure and scenario planning requires a high degree of accuracy in modeling the vehicle dynamics. This section highlights the calibration and validation techniques along with the microsimulator results. Previous studies have relied on vehicle counts, queue lengths at intersections, and vehicle speeds at loop detectors as ground truth data for calibration and validation \cite{technicalactivitiesdivisionDynamicTrafficAssignment2011}. 

In this work, we adopt a novel approach of calibration and validation using link-based speed data from the open-access Uber Movement project (\url{https://movement.uber.com/explore/san_francisco/speeds}). Uber Movement uses GPS data from Uber vehicles to calculate speeds at particular snapshots in time. The edge speeds are calculated using the GPS locations of vehicles and the map data of the street network. The Uber app records a vehicle's location information every 1 to 2 seconds, including latitude, longitude, speed, course, and timestamp of the GPS location ping. Uber uses map-matching with the latitude and longitude of a particular GPS ping to determine when a driver enters and exits an edge. The duration a vehicle spends on an edge is calculated as the time difference between when the driver enters the street segment and when the driver leaves that segment. The edge speed is then calculated as the length of the segment divided by the time taken to traverse. Uber does not disclose the volume data, but notes that they ensure the validity of the data by setting a minimum threshold for observations.

\subsection{Calibration}
\label{sec:calibration}
Traffic microsimulators require calibration to real-world data to adequately represent observed dynamics across a wide range of network structures and conditions~\cite{barceloDynamicNetworkSimulation2005}. In the IDM, parameters $a$, $b$, $T$, and $s_0$ are calibrated. The objective of the calibration process is to minimize the sum of the errors between every edge's speed from MANTA and the Uber data (L1 norm). This optimization problem is specified in~\Cref{eq:calibration},
\begin{equation}
    \min{\sum_{n=1}^N}{|\frac{\sum_{k=1}^K{v_{k,n}}}{K} - \overline{v}{_{u,n}}|}
    \label{eq:calibration}
\end{equation}
where $v_{k,n}$ is the calculated velocity of vehicle $k$ on edge $n$, $v_{u,n}$ is the average Uber velocity of edge $n$, $K$ is the number of cars on edge $n$, and $N$ is the number of edges that were successfully matched between Uber's street network and MANTA's street network. Expanding further in~\Cref{eq:calibration_full},
\begin{equation}
    \min_{a,b,T,s_0}{\sum_{n=1}^N}|{\frac{\sum_{k=1}^K{\dot{[a(1 - (\frac{v_{k,n}}{v_{0,n}})^\delta - (\frac{s_o + Tv + \frac{v\Delta v}{2\sqrt{ab}}}{s})^2]}t}}{K} - \overline{v}{_{u,n}}|}
    \label{eq:calibration_full}
\end{equation}
where $t$ is the timestep (set as .5 seconds), $a$ is the acceleration potential, $b$ is the braking potential, $T$ is time headway, and $s_0$ is the linear jam distance.

\begin{figure}
    \centering
    \includegraphics[width=.45\textwidth]{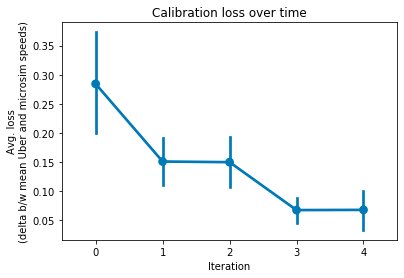}
    \caption{The calibration process: average mean difference between Uber and MANTA speeds over time}
    \label{fig:calibration}
\end{figure}

Given the highly nonlinear nature of the objective function, a numerical method is used to optimize the IDM simulation parameters. We constrain the acceleration and deceleration potential, $a$ and $b$, respectively, to $[1,10]$ meters per second squared, headway time $T$ to $[0.1,2]$ seconds, and linear jam distance $s_0$ to $[1.0,5.0]$ meters, and set the standard exponent of the IDM, $\delta$, to 4~\cite{treiberTrafficFlowDynamics2013}. A mini-batch gradient descent is then carried out across the entire simulation, with each iteration executing runs for 5 different sets of {$a$, $b$, $T$, and $s_0$}. We accumulated the sum of difference in speed between MANTA and Uber for all edges. The goal is to find the set $\{a, b, T, s_0\}$ that minimizes this sum of differences. The set that produces the lowest mean difference is chosen as the nominal vector for the next iteration. Each parameter is then perturbed by a value chosen randomly, sampled from a uniform distribution, from $[-1,1]$ at the next iteration. Such a large range is used in order to produce meaningful differences across sets within the next iteration. The perturbation range then decreases by an order of magnitude at every iteration (e.g., iteration 3 uses $[-.1,.1]$, iteration 4 uses $[-.01,.01]$, etc.). The calibration process converges once the mean difference decreases below a desired threshold of $.05$ miles per hour, considering runtime limitations. As shown in~\Cref{fig:calibration}, the calibration process converges after five iterations.

Since the loss function in~\Cref{eq:calibration} is non-convex across all of the calibrated parameters, this mini-batch gradient descent method will produce a local minimum and not necessarily a global minimum. When a successive order of magnitude decay, producing a larger perturbation range at every iteration, we observed that the gradient diverges, indicating a minima. Further research is required to improve the calibration method to approach global optimality.

\subsection{Validation}

\begin{figure}
    \centering
    \includegraphics[width=.45\textwidth]{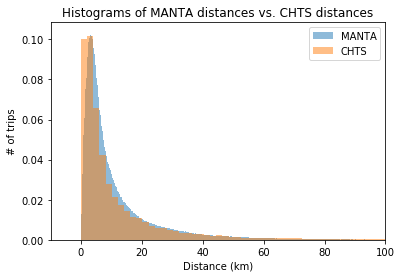}
    \caption{Comparison of trip lengths in MANTA versus California Household Travel Survey data. Median distance in MANTA is 6.46 km and in CHTS is 5.33 km.}
    \label{fig:distance}
\end{figure}

Validation is performed for both the routing algorithm and the traffic microsimulator. MANTA's routing algorithm is validated by comparing the routes against the California Household Travel Survey (CHTS) data for the SF Bay Area~\cite{nationalrenewableenergylaboratoryTransportationSecureData}.~\Cref{fig:distance} presents the distances traveled by each vehicle for both MANTA and CHTS. The distribution of distances is heavily right-skewed, suggesting that most trips are fewer than 25 km.  While CHTS data are sparse (69000 trips versus 3.2M trips in MANTA), we can still see similarities. MANTA estimates the mean distance traveled as 11.3 km, which is closer to 13.5 km in CHTS. Median values are 6.46 km and 5.33 km in MANTA and CHTS, respectively. The 75th percentile distances are also similar, at 13.6 km and 13.7 km for MANTA and CHTS, respectively. The modest differences between the MANTA and CHTS routing data may be attributed to stochastic error from random sampling of O and D location within the respective TAZs in MANTA simulation.

The validation of the traffic microsimulator involves comparing the MANTA outputs to Uber Movement distributions at specific timeslices. In particular, using Q2 Uber Movement data from 2019, we compare results of the MANTA simulations and the Uber data for 95,510 edges, or 17\%, of the total edges in the SF Bay Area network.

We enhanced the IDM to include varying maximum speed limits for each individual to better reflect real-world vehicular behavior. In the IDM, $v_0$ represents the free-flow velocity of a vehicle on an edge, typically the speed limit of each edge from OSM or from a standard convention. However, in order to mimic the variance of driving patterns across travelers, each driver's maximum possible speed per edge, $v_0$, is sampled from a Gaussian distribution centered around the edge's predetermined speed limit with a standard deviation of $2\sigma_s$, where $\sigma_s$ is the standard deviation of vehicle speeds, obtained from the Uber data, at each speed limit $s$. Every vehicle thus has a slightly different maximum allowable speed on each edge it traverses.

\begin{figure}
    \centering
    \begin{subfigure}{\linewidth}
        \includegraphics[width=\linewidth]{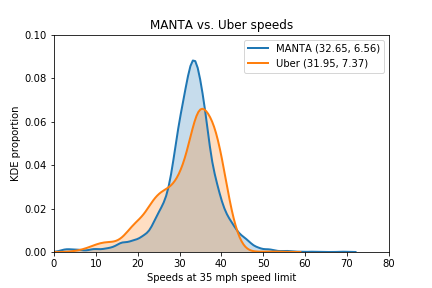}
        \caption{5 AM - 6 AM}
        \label{fig:kde_5to6_35}
    \end{subfigure}\\
    \begin{subfigure}{\linewidth}
    \includegraphics[width=\linewidth]{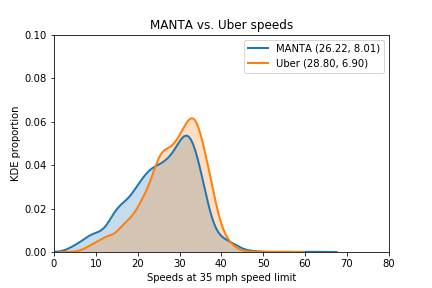}
    \caption{8 AM - 9 AM}
    \label{fig:kde_8to9_35}
    \end{subfigure}
    \caption{Kernel density plot comparing the MANTA and Uber distributions at 35 mph}
\end{figure}

We compare the distribution of speed on different edges between MANTA simulation and Uber data. For the simulation run between 5 AM - 12 PM, we investigate the difference in behavior at two different time periods: 5 AM - 6 AM, a less congested time period, and 8 AM - 9 AM, a more congested time period. Within each time period, we look at the speed distribution curves at different speed limits. For instance,~\Cref{fig:kde_5to6_35} and ~\Cref{fig:kde_8to9_35} show the speed distributions from MANTA on edges with 35 mph compared to the Uber Movement data on those same edges, at the representative time periods. As expected, both MANTA and Uber average speeds are higher between 5 AM and 6 AM (less congested) than those between 8 AM and 9 AM (more congested time period).

\begin{figure}
    \centering
    \begin{subfigure}{\linewidth}
        \includegraphics[width=\linewidth]{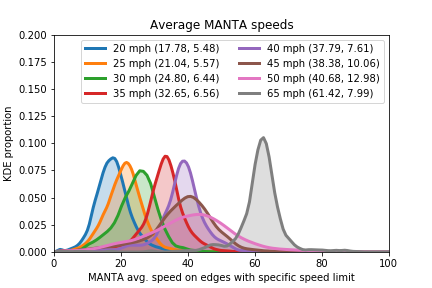}
    \end{subfigure}\\
    \begin{subfigure}{\linewidth}
        \includegraphics[width=\linewidth]{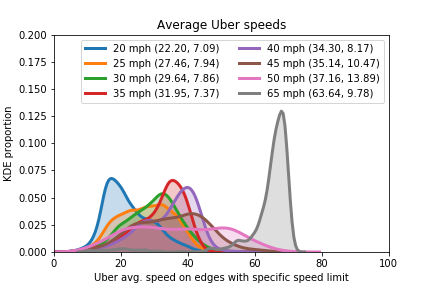}
    \end{subfigure}
    \caption{Average MANTA and Uber speeds across all speed limits [5 AM - 6 AM]. The means and standard deviations are shown in parentheses.}
    \label{fig:avg_microsim_vs_uber_5to6}
\end{figure}

\Cref{fig:avg_microsim_vs_uber_5to6} shows the average speeds of MANTA and Uber across all speed limits between 5 AM and 6 AM. At low-speed edges ($<$ 30 mph), MANTA simulation speeds are approximately 5 mph slower than Uber's real-world data. This indicates that the congestion effects are larger at lower speeds in MANTA. The Uber speeds also reflect that, in the real-world, many drivers tend to go above the speed limits more so on edges with lower speed limits than they do on edges with higher speed limits. For edges with speed limits above 30 mph, MANTA estimates may be higher or lower than the Uber estimates. This suggests that improvements can be made in both calibration and in modeling the individual behavior of drivers with respect to speed limits.

\begin{figure}
    \centering
   \begin{subfigure}{\linewidth}
        \includegraphics[width=\linewidth]{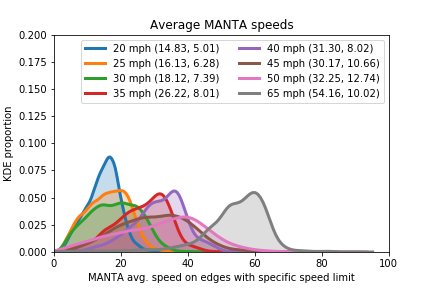}
    \end{subfigure}\\
    \begin{subfigure}{\linewidth}
        \includegraphics[width=\linewidth]{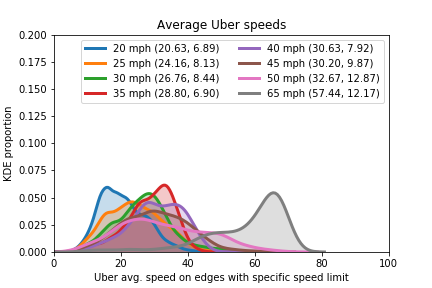}
    \end{subfigure}
    \caption{Average MANTA and Uber speeds across all speed limits [8 AM - 9 AM]. The means and standard deviations are shown in parentheses.}
    \label{fig:avg_microsim_vs_uber_8to9}
\end{figure}

\Cref{fig:avg_microsim_vs_uber_8to9} shows the distribution of speeds for the 8 AM - 9 AM timeframe. Unlike the less congested 5 AM to 6 AM timeframe, MANTA simulation speeds are equal to or slower than Uber's real-world data across all speed limits. This indicates that the IDM in MANTA may be overly sensitive to congestion effects.

Comparing the 5 AM - 6 AM timeslice with 8 AM - 9 AM in MANTA, the average speeds estimated in the early morning time period in general are higher by 3 to 9 mph across all speed limits, with the greater differences being on edges with higher speed limits. This intuitively suggests that roads with higher speed limits, such as highways,  see less traffic at the early morning hours, and thus vehicles can travel at higher speeds due to the lack of congestion and lack of stoppage. However, roads with lower speed limits do not allow for much higher speeds regardless of the time of the day. This is likely due to the presence of frequent intersections in the city. The Uber data across the two timeslices also reflect this difference.



\subsection{Red light / green light cases}
\label{sec:lights}
\begin{figure}
    \centering
    \includegraphics[width=0.49\textwidth]{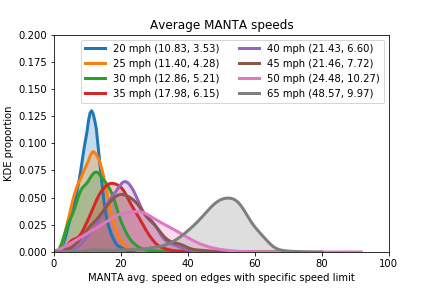}
    \caption{Average MANTA speeds across all speed limits [8 AM - 9 AM] in the red light case. The means and standard deviations are shown in parentheses.}
    \label{fig:velocities_red_light}
\end{figure}

In this study, we adopt a basic intersection model and consider two different conditions: where every node is either a flashing red light or a green light. In the flashing red light scenario, every vehicle is designed to stop at the intersection for 2 seconds before proceeding, similar to a stop sign, which can result in vehicles backing up and subsequent queue spillback. In the green light scenario, every vehicle can immediately access the intersection and proceed with its next move. \Cref{fig:velocities_red_light} shows the distribution of average speed across different speed limits. Between 5 AM - 6 AM, the average speed is 17.5 mph, while the speed decreases to 12.9 mph in the 8 AM - 9 AM timeslice. The reduction in speed between 8 AM and 9 AM suggests increased congestion, in comparison to the free-flowing traffic in the early morning between 5 AM - 6 AM. 

When every node is a green light, the average speed across all speed limits for the 5 AM - 6 AM time period is 24.5 mph. The average speed decreases to 17.8 mph for the 8 AM - 9 AM time period (see~\Cref{fig:avg_microsim_vs_uber_8to9}). The difference in speed limit between the early morning timeslice and the 8 AM - 9 AM peak hour timeslice in the green light case is 4.6 mph, while in the red light condition, it is 6.7 mph. The deltas between the two timeslices, as well as the absolute speeds, highlight notable differences in the traffic behavior between the two timeslices. Specifically, the average speeds in both timeslices under the red light condition is about 5 mph lower than the green light condition. Such low speeds are unsurprising given that every vehicle must stop and wait its turn in the intersection queue. Since the IDM parameters have been tuned to the real-world Uber data, which is better represented by the green light scenario, the red light scenario does not match the Uber data as closely as the green light scenario does.



\begin{figure}
    \centering
    \includegraphics[width=.45\textwidth]{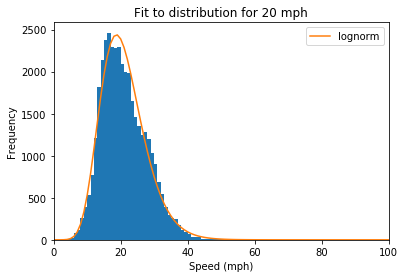}
    \caption{Fit to lognormal distribution for 20 mph speed limit in green light scenario (case 2)}
    \label{fig:lognormal_20_mph_green_light}
\end{figure}

\begin{figure}
    \centering
    \includegraphics[width=.45\textwidth]{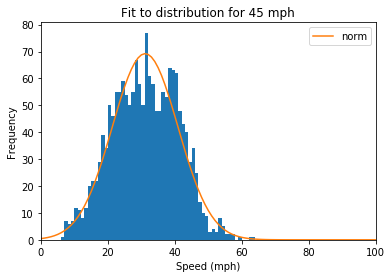}
    \caption{Fit to normal distribution for 45 mph speed limit in green light scenario (case 2)}
    \label{fig:normal_45_mph_green_light}
\end{figure}

Notably, in~\Cref{fig:avg_microsim_vs_uber_8to9}, the lower speed limits' distributions tend to be right-skewed, following a lognormal pattern, while the distributions at higher speed limits become more centered and follow a normal distribution. Snapshots of these phenomena are shown in~\Cref{fig:lognormal_20_mph_green_light} and~\Cref{fig:normal_45_mph_green_light}.

\section{Performance benchmarks}

This section describes the computational performance of the two core components of MANTA: routing and the microsimulator engine. 

\subsection{Routing performance}

\begin{figure}
    \centering
    \includegraphics[width=\linewidth]{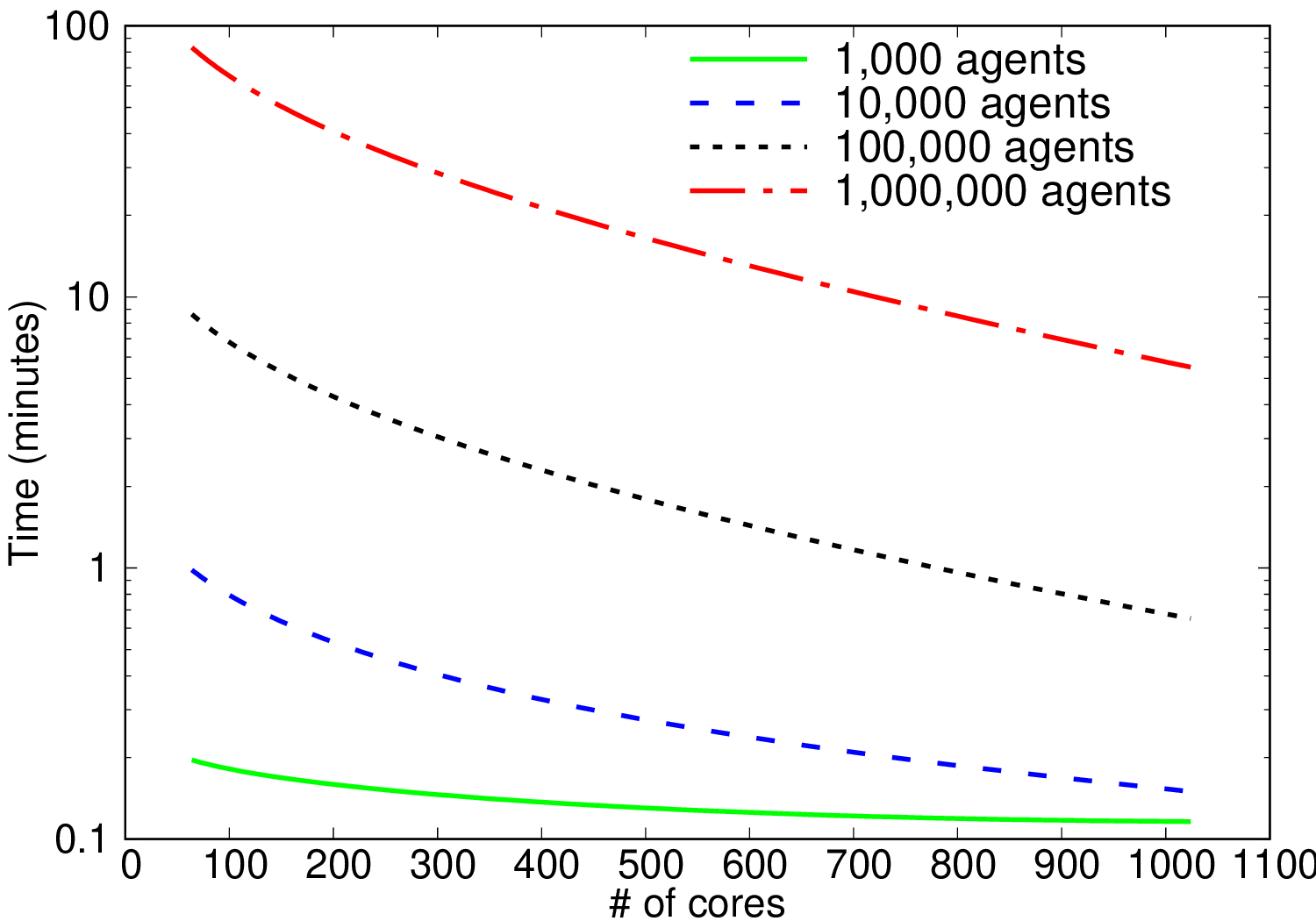}
    \caption{Time required to route agents using priority-queue Dijkstra algorithm for the SF Bay Area network on distributed computing environment (MPI + OpenMP) parallelization. Tests were run on 32 nodes with Intel Xeon Skylake 6142 processors.}
    \label{fig:time}
\end{figure}

\begin{figure}
    \centering
    \includegraphics[width=\linewidth]{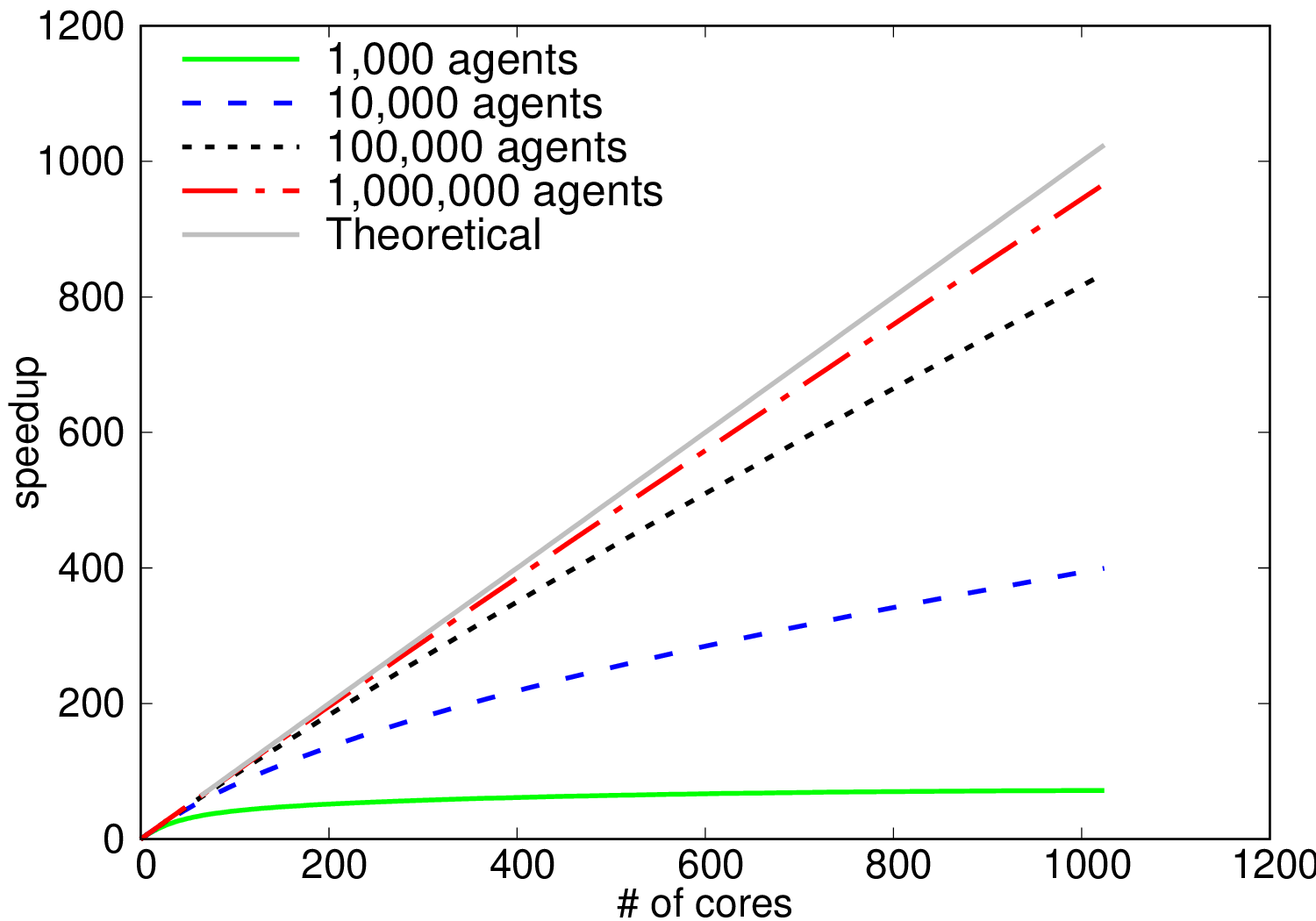}
    \caption{Speedup of priority-queue routing algorithm for the Bay-Area network on distributed computing environment (MPI + OpenMP) parallelization. Tests were run on 32 nodes with Intel Xeon Skylake 6142 processors.}
    \label{fig:speedup}
\end{figure}

In our network of approximately 225K nodes, 550K edges, and 3.2M OD pairs, the SSSP routing algorithm carries out the computation of all OD pairs' routes in approximately 62 minutes on a single node.~\Cref{fig:time} shows the time-required to run up to 1 million agents on a distributed compute cluster utilizing both MPI and OpenMP parallelization schemes.~\Cref{fig:speedup} shows that the strong scaling results of the routing algorithm matches the theoretical scaling up to 1024 cores for routing 1 million agents. In comparison to existing routing algorithms, such as the heuristic-based Ligra~\cite{shun2013ligra} and iGraph~\cite{csardi2006igraph}, the priority-queue based Dijkstra is 2.2\% and 55\% faster, respectively, on a single node. The priority-queue Dijkstra algorithm also has higher effective CPU usage of 94.1\% with an average RAM usage of 4.81 GB.

\subsection{Microsimulator performance}

The computational performance of the MANTA simulator is compared with Simulation for Urban Mobility (SUMO) and JDEQSIM, a parallelized alternative available in MATSim, two well-known open-source simulators in transportation. The simulation of the SF Bay Area network and the demand between 5 AM - 12 PM are used for the comparison exercise. SUMO offers two options to build the network: one that contains internal links or lanes within intersections, and one that does not contain internal links~\cite{dlr46740}. Considering MANTA's simplified intersection model, the SUMO model without internal links is the most appropriate comparison. The SUMO model with internal links is also included for completeness.

~\Cref{tab:runtimes} shows the runtime comparison of MANTA against SUMO and JDEQSIM. The table also indicates when the results are linearly extrapolated, due to the inability to complete simulations in a reasonable time. Extrapolating the simulation runtime linearly, MANTA performs nearly 27000x faster than SUMO. MANTA carried out the full microscopic simulation of 3.2M trips at .5 s timesteps in 4.6 minutes, while SUMO's simulator is estimated to take nearly 87 days, linearly extrapolated from the initial run of 194 minutes for 5000 trips. SUMO also has a mesoscopic simulator, which requires approximately 29 hours (1740 minutes) for the SF Bay Area simulation.


A primary reason for such a dramatic difference in runtimes is that typically SUMO uses a traffic assignment model for routing. When the routes are fixed, as in this example, SUMO sees undesired jamming, as many roads are not filled to their capacities while other roads are filled excessively. The resulting congestion increases the simulation time in SUMO to achieve equilibrium. Unlike SUMO, MANTA is a dynamic model and does not perform equilibrium traffic assignment. In other words, MANTA does not minimize the total travel time of the entire system, but instead assumes that each driver will take the shortest route based on distance. Notably, SUMO's microsimulation does not support parallelization; only the routing algorithm is parallelized, which is not germane for this comparison.

JDEQSIM is a discrete event-based mesoscopic simulator that uses event handling to communicate every person's activity to the rest of the Behavior, Energy, and Autonomy Modeling (BEAM) platform\cite{waraichPerformanceImprovementsLargeScale2015}. The event handler manages billions of activities and events (specifically when vehicles enter and exit edges), which produces a significant overhead in the generation and synchronization of events across the threads. MANTA, on the other hand, is a time-based simulator that does not have overhead from the constant generation of events. In addition, the texture mapping of MANTA is optimized for fast GPU array manipulation, which yields significant speedup compared to the CPU implementation in JDEQSIM~\cite{waraichPerformanceImprovementsLargeScale2015}. 

\begin{figure}
    \centering
    \includegraphics[width=.45\textwidth]{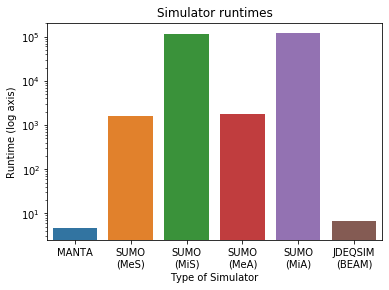}
    \caption{Simulator runtimes (log scale y-axis) across different simulators. MANTA performs slightly better than the parallelized mesoscopic JDEQSIM and is on the same order of magnitude. MANTA performs significantly better than the mesoscopic version of SUMO with either the simplified (MeS) or advanced intersection modeling (MeA). The microscopic version of SUMO with simplified intersections (MiS) and advanced intersections (MiA) could not be run completely, and thus times were linearly extrapolated, reflecting that it would take tens of days to complete.}
    \label{fig:sim_runtime}
\end{figure}

~\Cref{fig:sim_runtime} shows the comparison of runtimes between JDEQSIM and MANTA. The JDEQSIM runtime is approximately 6.6 minutes, on average over 50 runs, and is comparable to MANTA's runtime of 4.6 minutes. The GPU parallelized traffic microsimulation in MANTA is 43\% faster than aggregated simulators such as JDEQSIM. In comparison to the SUMO microsimulation, MANTA is several orders of magnitude faster. Considering the finer level of behavioral granularity achieved by MANTA at the runtime of the mesoscopic JDEQSIM, these results clearly demonstrate the applicability of MANTA for metropolitan-scale traffic microsimulations.

Other parallel microsimulators exist as well, including ~\cite{chanMobilitiScalableTransportation2018, barceloParallelizationMicroscopicTraffic1998, nagelParallelImplementationTRANSIMS2001}, but they either require expensive supercomputing facilities or carry out simulations on smaller networks with longer computation times.

\begin{table}
    \centering
    \begin{tabular}{l c c}
    \toprule
     \textbf{Simulator} & \textbf{Time (mins)} & \textbf{Type}\\
     \midrule
     MANTA & $4.6$ & Full\\
     SUMO meso simplified (MeS) & $1620$ & Full\\  
     SUMO micro simplified (MiS) & $114858$ & Lin. extrap.\\ 
     SUMO meso advanced (MeA) & $1740$ & Full\\
     SUMO micro advanced (MiA) & $123500$ & Lin. extrap.\\
     JDEQSIM & $6.6$ & Full\\
     \bottomrule
    \end{tabular}
    \label{tab:runtimes}
    \caption{MANTA's runtimes compared to SUMO and JDEQSIM. Full implies that the entire simulation was able to complete. Lin. extrap. implies that only part of the simulation was able to complete and the full time was linearly extrapolated from this preliminary time.}
\end{table}

\section{Limitations}

The traffic microsimulation in MANTA achieves significant advances in computational performance using metropolitan-scale networks and demand, but important limitations remain. The first limitation is the use of simplified intersection modeling. A more accurate intersection modeling will produce precise travel times and a better representation of the vehicle dynamics.

The second limitation is the demand profile. This work uses a synthetic Bay Area MTC 2017 travel model that represents the daily demand in five large time blocks and carries out a static traffic assignment. A more realistic model could integrate a dynamic travel demand model, such as ActivitySim, with MANTA. 

The modular structure of MANTA offers the ability to vary different components of the network analysis, such as routing and vehicular dynamics. MANTA currently can accommodate different routing algorithms, such as Dijkstra, A*, and Contraction Hierarchy. In addition, while MANTA currently uses the Intelligent Driver Model, it has the functionality to leverage other driver models. Incorporating dynamic routing, where the edge weights are based on travel times on the edge rather than the length of the edge, will improve the predictive accuracy of near-real-time simulations, such as evacuations. 


\section{Conclusions}

This paper presents a novel traffic microsimulator, MANTA, that addresses the challenges of accurate traffic microsimulation at the metropolitan-scale. MANTA is highly efficient and is capable of simulating real-world traffic demand with a fine level of granularity on very large-scale networks. The runtime efficiency of MANTA is achieved by efficiently coupling a distributed CPU-parallelized routing algorithm and a massively parallelized GPU simulation that utilizes a novel traffic atlas to map the spatial distribution of vehicles as contiguous bytes in memory. The capability of MANTA is demonstrated by simulating a typical morning workday of the nine-county SF Bay Area network with 550K edges and 225K nodes, and approximately 3.2M OD pairs. The routing calculations are completed in 62 minutes, and a simulation of 7 hours from 5 AM to 12 PM with .5 second timesteps is completed in 4.6 minutes. This is several orders of magnitude faster than the state of the art microsimulators with similar hardware. Achieving compelling performance in both efficiency and accuracy, MANTA offers significant potential for fast scenario planning in both short- and long-term applications in metropolitan and metropolitan-scale analysis.

\section{Acknowledgements}

This report and the work described were sponsored by the U.S. Department of Energy (DOE) Vehicle Technologies Office (VTO) under the Systems and Modeling for Accelerated Research in Transportation (SMART) Mobility Laboratory Consortium, an initiative of the Energy Efficient Mobility Systems (EEMS) Program. The following DOE Office of Energy Efficiency and Renewable Energy (EERE) managers played important roles in establishing the project concept, advancing implementation, and providing ongoing guidance: David Anderson, Rachael Nealer, and Erin Boyd as well as Prasad Gupte. This work was funded by the U.S. Department of Energy Vehicle Technologies Office under Lawrence Berkeley National Laboratory Contract No. DE-AC02-05CH11231.

The authors would like to give a special thanks to Kenichi Soga, Bingyu Zhao, and the cb-cities research group at the University of California, Berkeley and the University of Cambridge; Rashid Waraich, Artavazd Balayan, and the BEAM project team at Lawrence Berkeley National Laboratory; and the SUMO open-source team for remote simulation support.

\section{Appendix}
MANTA is an open-source research code distributed under BSD 3-clause license and is available at~\url{https://github.com/UDST/manta}.

\bibliographystyle{ieeetr}
\bibliography{030121_microsim.bib}

\end{document}